RESEARCH ARTICLE

# Should there be a teacher in-the-loop? A study of generative AI-personalized tasks middle school

Candace Walkington[1] · Mingyu Feng[2] · Tiffini Pruitt-Britton[2] · Theodora Beauchamp[1] · Andrew Lan[3]

Received: 6 March 2026 / Accepted: 23 October 2026

**Abstract**
Large language models can be used to design instructional tasks to align to students' needs. These generative AI models can create tasks that correspond to students' out-of-school interests in areas like sports or music, which may enhance students' interest in learning. However, little is known about teachers' processes for creating or modifying tasks with generative AI, despite commercial AI companies' claims that they will save teachers time. In the present study, 7 middle school mathematics teachers used ChatGPT to create personalized versions of problems in their curriculum to correspond to their students' interests. The prompting moves teachers made and their students' open-ended feedback on personalized problems were examined through thematic analysis. Teachers' efficiency when creating problems and the ratings of their 7th grade students who received personalized assignments were examined. Results from this sample suggest that having a teacher-in-the-loop may result in generative AI-enhanced personalization being enacted at a broad grain size, whereas students often expressed preference for a smaller grain-size where they receive specific popular culture references that interest them. Teachers used prompting to adjust popular culture references and address issues with the depth, ethics, or realism of the problems generated, in some cases showing epistemic agency. The article closes with an exploratory, design-focused look at scalability trade-offs with teacher-facing personalization, patterns of teacher prompting practices, and opportunities for teachers to exercise epistemic and ethical monitoring of personalization outputs.

✉ Candace Walkington
cwalkington@smu.edu

[1] Department of Teaching and Learning, Southern Methodist University, 6401 Airline Rd, Dallas, TX 75025, USA

[2] Learning and Technology Division, WestEd, San Francisco, USA

[3] Manning College of Information and Computer Sciences, University of Massachusetts Amherst, Amherst, USA

## Introduction

Personalizing instruction to the needs, goals, and interests of students has been a goal in education, framed as a “grand challenge” of the 21 st century (National Academy of Engineering, 2008). The rise of generative AI (GenAI) in the 2020 s allows new ways to meet this challenge, given GenAI’s ability to adapt tasks to learners’ characteristics. Large Language Models (LLMs) like ChatGPT are one type of GenAI that can recognize and generate texts using machine learning based on large amounts of training data. These technologies are scaling in education, with estimates that 60% of teachers were using GenAI during the 2024/2025 school year (Gallup and Walton Family Foundation, 2025). One application of GenAI for education is *context personalization*—the tailoring of educational tasks to students’ interests (Cordova & Lepper, 1996). A recent meta-analysis suggests that context personalization improves both interest and learning (Lin et al., 2024). However, a question remains: To what extent should teachers be in-the-loop during GenAI-enhanced instruction?

When considering context personalization with AI, different models have emerged. On one extreme, the teacher has agency to create tasks tailored to their students’ interests, experiences, and goals, based on their professional knowledge and the assistance of GenAI (e.g., Rutherford et al., 2025). This approach allows the teacher to be fully in control of their students’ learning experience, facilitates the teacher’s continued professional learning, and better connects teachers to their students. However, it may also be time-consuming for teachers, although most studies do not address, this issue directly. Further, the teacher cannot reasonably make a different version of every educational task for every single student and must choose which student interests to highlight. On the other extreme, the teacher is taken out-of-the-loop (e.g., Khan Academy, n.d.; see Norberg et al., 2025 for a related approach). In this approach, students can input something very specific—like Taylor Swift or Roblox—and every student can have their educational tasks adapted. An example of this approach is Khan Academy’s (n.d.) GenAI tutor, which can remember students’ interests and personalize tasks. These tasks can have an extremely small grain size. There are also possibilities for hybrid approaches that allow for different distributions of generative work, evaluative control, and decision authority between teachers and AI. A process might be designed where teachers review problems the AI generated based on student inputs, or where teachers offer significant model instructions, training examples, and feedback for fine-tuning AI generation systems that face students. The tradeoffs between these approaches can be motivated based on theories of personalization (Walkington & Bernacki, 2014), as these approaches may differ in the depth and authenticity of the problems students receive, the grain size or the precise fit of the problems to students’ interests, and issues of ownership of the personalized tasks.

The present study examines teacher-in-the-loop context personalization. Rather than treating personalization as a system-level feature, this study conceptualizes it as an AI-mediated design process in which teachers retained the possibility for epistemic authority (Wu et al., 2025). Analyses reveal micro-level decisions through which issues of depth, grain size, and ownership are negotiated by teachers, which can reveal instances of teacher epistemic agency. Seven middle school teachers created mathematics problems personalized to their students’ interests using GPT-4o. The teachers began with current problems in their *Illustrative Mathematics* (IM) curriculum and then modified problems to better match students’ interest areas. IM is a free, open-source mathematics curriculum used in the United States with materials for grades kindergarten through 12th grade. These problems were subsequently assigned to their students ($n = 521$). Analyses examine the prompts teachers used to design these personalized tasks, their efficiency when designing with LLMs, and their students' reactions to these tasks. This study explores teaching situations within an authentic curricular implementation, integrating multiple student- and teacher-level data sources, with a unique focus on teacher time investment and prompting behavior when working with LLMs to personalize tasks. It provides an exploratory account of tensions among grain size, scalability, and epistemic agency in teacher-in-the-loop personalization.

## Theoretical framework and literature review

### Depth, grain size, and ownership

*Context personalization* is defined as adapting instruction to student interests (Cordova & Lepper, 1996), and this study focuses on interests relating to popular culture areas like sports and social networking. Personalization has been shown to trigger students' *situational interest* (a state of attention and heightened affect and engagement) in a learning task, which can become maintained over time (Bernacki & Walkington, 2018). Personalization can increase immediate performance, learning efficiency, and long-term learning and interest (Bernacki & Walkington, 2018; Lin et al., 2024; Walkington, 2013). Developing personalized task variations is time-intensive, particularly when tasks are to represent students' actual practices in daily life. The process of teachers personalizing a task may involve iterative refinement and teacher agency—where teachers need time and space to engage in meaningful rounds of personalization with appropriate supports, critically evaluating what they produce. However, some tasks may be difficult to personalize because of their structural characteristics and the difficulty of connecting certain interest areas to specific academic concepts, leading to the possibility of diminishing returns for increased investments of teacher effort in personalization.

Walkington and Bernacki (2014) identified three considerations for personalization design, which form the core of the theoretical framework. *Depth* is the degree to which the personalization reflects how students *actually use* the content while pursuing their interests. For example, a problem about how a student's number of followers changes over time on Instagram may relate to their actual quantitative experiences with the platform, while a generic problem about Instagram revenue may be less aligned. Research shows that shallow approaches to personalization that only draw on surface-level characteristics of students' interest may achieve effects on situational interest but not on performance or learning (Høgheim & Reber, 2015) or obtain no effect (Kosh, 2016), but results vary (Cordova & Lepper, 1996). Depth can be reflected in the realism of the personalized scenarios, or their correspondence with how things happen in the world—whether the event described in the personalized task would occur, whether the numbers are realistic and appropriate, and whether the quantities are those that would be used and measured (Palm, 2008). When teachers personalize tasks with LLMs, prompts may show the teachers considering depth when the teacher makes corrections to preserve the problem's realism, or when the teacher works to make the LLM output more closely match the students' actual interests and practices, so that it will activate relevant prior knowledge. When students give feedback on problems, they may comment on depth when they note a problem is not realistic or is not faithful to things they would do in their lives.

*Grain size* is the degree to which learners' *particular* interests are reflected in the learning task (Walkington & Bernacki, 2014). Indeed, tasks can be personalized at the level of an individual student's interests, at the level of a class's interests, or at the level of common interests to an entire age group. For example, a student interested in baseball might be placed into a generic "sports" category where they get tasks about swimming. The recent meta-analysis on context personalization suggested that smaller grain-size approaches are best for enhancing students' interest, with an estimated effect size of $g = 0.89$ (Lin et al., 2024). In the present study, personalization occurred at the relatively broad grain size of the interests of an entire class (as perceived by the teacher) rather than at the grain size of a particular student. However, there were different ways in which the problems could be designed that also impact their grain size. Teachers can write problems that have *specific* popular culture references (e.g., going to a Taylor Swift concert) or *general* popular culture references (going to a generic music event). The latter may have lower appeal to many students, while the former may have higher appeal to a few students. When teachers personalize tasks with LLMs, prompts may show the teachers considering grain size when they ask the LLM to change more general references in the problem (e.g., a music show) or references they do not think would resonate with their students (e.g., a show of a classical music artist) to more specific references that would resonate with their students (e.g., the show of a country music artist they know their students like). Prompts may also show considerations of grain size when the teacher specifies other characteristics of their students—like their age group, geographic location, or typical first names. When students give feedback on problems, they may comment on grain size when they express liking or disliking for particular interest references in the problems, as they relate to their preferences.

Finally, *ownership* of personalization concerns who controls the personalized connections being made or the specific student interests being highlighted. Teacher control can create opportunities for teachers to exercise epistemic agency (Wu et al., 2025), using their knowledge to apply criteria to the evaluation of AI outputs. Control over personalization has traditionally been distributed among teachers, students, and curriculum developers or researchers (Walkington & Bernacki, 2014). However, with the rise of LLMs, there is a new element of ownership that has been introduced—LLMs can be given more or less control over the personalized content development. The amount of control the LLMs are given is determined by the teacher in teacher-in-the-loop models—the teacher chooses how they want to use the LLM, for what purpose, and to what degree, resulting in higher or lower levels of control for the LLM. When teachers personalize tasks with LLMs, prompts may show considerations of ownership when the teacher gives the LLM specific student interests themselves based on their professional knowledge, versus letting the LLM use its processes to determine likely student interests. Adding students' first names to problems and incorporating highly-specific interest topics that the student would know came from them, may also enhance their perceptions of ownership.

Considerations of depth, grain size, and ownership can be overlapping in important ways—making a problem more specific to particular student interests may also allow for the student to make deeper quantitative connections to the problem due to enhanced familiarity. Making a problem very fine-grained might also signal to students that the problem was personalized with them in mind, activating ownership. And giving the LLM ownership of choosing interest references for problems might enhance depth and grain size if the teacher is very unfamiliar with the interest area and the students use common interest practices for their age group but may compromise them if the students' interests and experiences do not match typical LLM responses.

## Ethical issues

There are important ethical issues to consider as teachers personalize tasks with LLMs. Research suggests that teachers can struggle to find mistakes in problems that LLMs pose, challenging the usability of these systems (Kim et al., 2026; Kwon & Ko, 2024). LLMs sometimes struggle with understanding the use of different quantities in word problems, especially when real-world knowledge is required (Srivatsa & Kochmar, 2024). Further, research has shown that ChatGPT can create developmentally inappropriate activities such as a middle school scenario about someone losing 5 pounds per month in a weight loss program (Sawyer & Aga, 2024). LLMs' lack of authentic connections to learners' lived experiences can "demonstrate a dangerous surface-level approach to culturally relevant pedagogy" (Gómez Marchant & Hardison, 2024, p. 3), indicating issues with depth and authenticity. Further, generative AI shows cultural, racial, and gender biases, which could be reflected in the mathematical tasks that it generates (e.g., UNESCO, 2024; Walkington et al., 2026a). LLMs are trained on large corpora of existing math problems, which textbook analyses have shown to be culturally biased across dimensions like names and images of characters, gender roles, and the kinds of activities being depicted (Pruitt-Britton & Walkington, 2022). Finally, using LLMs to write or rewrite mathematical tasks can result in tasks that are less readable (Huang et al., 2024; Luik et al., 2025), making these problems potentially less accessible for English-Language Learners and students who struggle with reading.

## Teacher instructional planning with LLMs

Research suggests that when using GenAI to engage in lesson planning, teachers use GenAI to enhance their own conceptual understanding of concepts, to refine and adjust lesson components, including both the academic content of the lesson and the instructional methods and scaffolds (Kim, 2026). Challenges include the potential for long, iterative prompting sequences, checking for reliability, and aligning outputs with teachers' contextual knowledge of student learning needs and expectations (Kim, 2026). Indeed, teachers notice the shortcomings of AI-generated content, such as lack of details or supporting materials provided, errors in the output, and the need for precise prompts and finetuning (Rutherford et al., 2025; Segal & Biton, 2024). Teachers' integration of GenAI outputs into their lesson plans varies, with interaction patterns ranging from directly using AI outputs, to using them with revisions, to critical analysis with reprompting and elaboration, to rejecting AI outputs (Choi et al., 2026). These usages can vary based on teacher experience (Choi et al., 2026) and AI proficiency. GenAI can act as a cognitive scaffolding tool in lesson planning, where its responses are a starting point for idea development and critical discussion, but this type of usage can be less common than basic operational tasks and can diminish over time as lessons are developed (Sun & Huang, 2025).

When considering the professional learning experiences that teachers should receive on GenAI integration, research and theory suggest that core competencies for teachers include technical fluency, pedagogical integration, critical evaluation, ethical reasoning, and collaborative practice (Shenoy & Saarela, 2026). Teachers can progress through different developmental stages as they learn to use AI, including exploration with GenAI tools, integration through AI-assisted lesson design, critical reflection of AI use, and peer exchange where there is human mentorship around AI use (Shenoy & Saarela, 2026). Promising professional-learning approaches combine hands-on experimentation with pedagogical and ethical framing, including explicit attention to bias and critical evaluation of AI-generated outputs (Tran et al., 2026). Sustained co-design approaches additionally provide structured peer feedback and repeated opportunities to design, evaluate, and revise instructional materials (Delaney et al., 2025; Segal & Klemer, 2025). Such opportunities to engage in these practices while designing real mathematical tasks can allow for qualitative gains to be seen in mathematics teachers' technological pedagogical content knowledge (Segal & Klemer, 2025).

One way in which instructional planning with LLMs has been integrated into math teachers' practices is through problem-posing activities where teachers write math problems with AI. Problem-posing activities with ChatGPT involve teachers using prompts to enhance problem clarity, increase problem challenge, include real-world applications, identify potential student difficulties, and to better focus the problem's mathematical topic (Biton & Segal, 2025). Teachers note that sometimes LLMs give unrealistic problems or give only shallow modifications for learners with different characteristics (Beauchamp & Walkington, 2024). When creating tasks with ChatGPT to support students' motivation, teachers acknowledge that it can be time-consuming to design with ChatGPT and they varied in the depth of their engagement with students' out-of-school interests and in whether they relied on interest data students reported or their own professional knowledge (Rutherford et al., 2025).

When teachers work with GenAI, they can express *epistemic agency*, which refers to a teacher's current tendency to actively shape the knowledge and practices in their professional communities (Wu et al., 2025). Here that agency is considered relational and temporally emergent, developing over time through teacher interactions with various elements of their social and material systems, including GenAI (Cathcart et al., 2026). Epistemic agency is evident when teachers explicitly use professional, disciplinary, pedagogical, or contextual knowledge to establish criteria for, evaluate, and modify AI-generated representations. General teacher decision-making, prompt refinement, and quality control can overlap with epistemic agency, but they do not constitute it by themselves; they provide evidence of epistemic agency when the teacher invokes and acts on a knowledge-related warrant.

In this study, we use *epistemic monitoring* to refer specifically to the evaluative component of epistemic agency, where teachers examine AI-generated mathematical and contextual representations against knowledge-related criteria to decide whether they should be accepted, investigated further, revised, or rejected. By knowledge-related criteria, we mean criteria like the validity of the mathematics, the realism of the scenario, and the authenticity of the scenario with respect to students' experiences, which can correspond to teachers' epistemic ideals (Barzilai & Chinn, 2024). Such criteria are contrasted with other kinds of actions we would not consider to be epistemic monitoring, such as organizational, stylistic, or preference-based modifications to AI outputs, which may not always reflect the teachers' explicit monitoring relating to knowledge and knowing. We then distinguish *ethical monitoring* as the teachers' evaluations related to potential harm, bias, exclusion, and developmental appropriateness, which is a key kind of monitoring when considering AI tools and their impact on children.

Teachers can have different epistemic stances when using GenAI, which are defined as the dynamic position they occupy relating to acquiring and validating the knowledge they receive from AI against specific criteria they determine. Stances can vary from being absolutist, where AI outputs are accepted uncritically, to multiplist, where teachers recognize that there are multiple viewpoints and may inquire into reasons behind different methods and AI outputs, to evaluativist, where outputs are critically evaluated for their suitability, integrating multiple perspectives including AI (Wu et al., 2025). Each of these stances has implications for the ways in which teachers interact with GenAI to personalize tasks to students' interests.

### Research questions

The present study addresses several gaps in the literature. First, much of the research on how teachers use GenAI to create instructional tasks has been conducted with pre-service teachers. In such studies, teachers do not have the opportunity to get feedback from students and iterate. This iteration can allow teachers to proactively address concerns with depth and grain size, engaging in epistemic monitoring. Further, in current studies teachers are almost always creating new mathematical tasks rather than modifying existing tasks—while the latter might be more realistic for teachers who use prescribed curricula. In addition, little research has explored teachers posing personalized problems with LLMs, instead focusing on either teachers posing any type of math problem or student-facing personalization where teachers are not in-the-loop. These scenarios have significantly different considerations related to depth, grain size, and ownership, as well as possibilities for teachers to exercise epistemic agency. Existing research often relies solely on teacher perceptions, rather than examining teachers' prompting practices and time measures. This can reveal important information about how teachers are responding to issues of depth, grain size, and ownership as they collaborate with AI in ways that may involve knowledge-based ethical and epistemic monitoring. And finally, few empirical studies precisely measuring teachers' time spent collaborating with AI have been conducted. The research questions are:

1) What kinds of LLM prompts do teachers use to modify existing mathematics problems to be personalized to students' interests, and how do they relate to depth, grain size, and ownership?
2) What reasons do these teachers' students give for finding the personalized problems interesting or uninteresting and what suggestions do they have for improving the problems, relating to depth, grain size, and ownership?
3) How efficient is personalizing existing problems using an LLM for teachers, and how is the teachers' prompting associated with how the students rate those problems for their interestingness and their closeness to students' interests?

## Method

### Sample

The sample included 7 teachers (mean teaching experience = 11.9 years) posing 143 personalized problems, which were then given to 7th grade students in ASSISTments (Heffernan & Heffernan, 2014), a platform for math homework that includes the IM curriculum. Teachers taught 7th grade mathematics in 6 different US schools; 2 teachers identified as male and 5 as female. Six identified as White and 1 identified as multi-racial (Black and White). The schools were geographically diverse (East Coast, West Coast, South, Midwest) and served an average of 30% students who were economically disadvantaged (range of 3 to 55%) and 49% students from racially/ethnically marginalized groups (range of 13 to 99%). Five of the 7 teachers had experience using both ASSISTments and IM. Teachers were recruited with email advertisements via ASSISTments and were paid for their study participation.

This study was approved under IRB #24-073 at Southern Methodist University. Parents could either (1) opt-out of research participation for their child, (2) not opt-out, and allow de-identified ASSISTments log data (with 1–5 problem ratings) to be collected from their child, or (3) opt-in to demographic and open-ended feedback data being collected from their child in addition to ASSISTments log data. The 7 teachers taught 521 students whose parents did not opt-out and 13 students whose parents did opt-out. 197 students had the additional layer of opt-in that allowed us to collect demographic information in addition to ASSISTments log data. Seventy-six students (39%) identified as female, 88 (45%) as male, 6 (3%) as Other or preferred not to respond, and 27 (14%) missing. For race/ethnicity, 121 (61%) identified as White, 25 (13%) as Hispanic, 7 (4%) as Black, 6 (3%) as Asian, and 10 (5%) as Other, with 28 (14%) missing. The teachers varied in their limited prior reported usage of GenAI—from not using it at all, to using it for communication with families, to occasionally using it to generate problems and worksheets.

Because this was an exploratory, design-focused study, the adequacy of the sample size was considered in relation to the richness and repetition observed in the data. The seven teachers contributed 143 problem-design episodes and 558 prompts, providing a substantial corpus for characterizing prompting practices, illustrating recurring design tensions, and generating hypotheses about teacher-in-the-loop personalization. The student sample also consisted of a large corpus of open-ended student feedback responses, which has been uncommon in work on personalization and problem-posing. However, the study was not intended to estimate the prevalence of these practices among mathematics teachers or students generally.

## Procedure

Teachers participated in a 2-h training where they learned about research on personalization, explored how to elicit their students' out-of-school interests, discussed ethical issues of bias, readability, and cultural relevance when personalizing problems, and practiced using ChatGPT to personalize problems. During the training, the teachers were encouraged to collect information about their students' interests for their participation in the study, but their exact processes were not tracked to simplify issues with parental consent. During this training, teachers and researchers discussed different methods they could use to collect their students' interests (see presentation slide, Appendix A, Fig. A1). Teachers were given a student interest interview protocol (Appendix A, Fig. A2) and a student interest survey (Appendix A, Fig. A3), and each strategy for assessing student interests was discussed. Teachers then selected up to 8 problems from IM Unit 4 (Proportional Relationships and Percentages). The teachers made either 1 or 2 personalized versions of each problem, using consistent interest areas (e.g., sports or food), using ChatGPT (GPT-4o) to write personalized versions of each problem while being recorded by Zoom. The teachers were cognizant during the process of creating problems that were relevant to different genders. Appendix A (Table A1) shows the interest topics teachers used for each assignment. Teachers were asked to consider heterogeneity within their classes rather than assume that one topic would appeal to all students of a particular gender. Most often two versions of each problem were created, and students could choose between them. This reduced the chances that students would not have an alternate choice if there was a gender association with a particular topic that a student finds problematic.

Final problems were turned into problem set(s) in ASSISTments. When there were multiple personalized versions of the same problem (e.g., one about sports and one about food), students were enabled to choose a version. Students completed the assignments for classwork or homework, and then the researchers and the teacher engaged in a reflection session together to go over student feedback and ratings. During these sessions, one of the tasks was for the teachers to examine students' open-ended feedback to get ideas for what interest topics to use in their next personalized assignment. A summary of the interest topics that were uncovered from student feedback for discussion during the reflection sessions is given in Appendix A (Table A2). Also during these sessions, teachers would reflect on the quantitative ratings of their problems, on readability metrics calculated relating to their problems, and on researcher comments relating to authenticity and bias in their problems. Teachers were asked to take notes, such that the reflections could influence their next problem-writing session. They might plan to make improvements when writing their next problem set based on these sessions, focusing on problem elements like the interest topics they selected, the readability of the problems they tended to include, or the authenticity or level of bias and how they noticed these issues. These reflection sessions were video-recorded and transcribed. They were used as a source of contextual background data to triangulate or allow investigation into issues that arose in the other data sources; they were not themselves coded in any formal way.

This process of creating and assigning personalized problem sets was then repeated for IM Unit 6 (Expressions, Equations, and Inequalities), with another reflection session taking place after Unit 6. Teachers posed an average of 20.4 problems ($SD$ = 8.1) across the two sessions. Appendix B shows IM problems before and after personalization, and Appendix C shows teachers' interactions with ChatGPT.

## Measures

### Coding of ChatGPT logs

Teachers' interaction logs with ChatGPT were analyzed for how many prompts they used and the time they spent interacting with ChatGPT. Then the type of prompt was coded using inductive coding. For 19 of the 143 problems the teacher was off-camera or not sharing their screen, so no time data was recorded.

### Video recordings of teachers

Teachers' speech as they generated each problem was transcribed and then divided into chunks to match the problems it corresponded to (see Appendix C). For example, a teacher might make a verbal comment saying that her students struggle with fractions before asking ChatGPT to change the numbers in a problem. These statements allow us to understand the reasons behind the prompting moves that teachers made that might have been invisible otherwise. Teachers were instructed to explain their actions in ChatGPT to 2 researchers who were present while they worked on personalizing. They were often also asked to give a verbal evaluation of what they thought of the final problem if they hadn't engaged in significant verbal comments while writing it. Researchers proactively asked teachers questions if they felt the reasons behind the prompting moves were not clear. There was an average of 210 words exchanged per personalized problem teachers wrote (SD = 204, range 0–1057). Appendix C gives multiple examples of what these conversations looked like. The video footage for one session was missing, so only ChatGPT logs were used.

### Student closed-ended survey ratings

For 53 of the 143 personalized problems, all students received two items: (1) "On a scale of 1–5, how interesting did you find this last problem?" and (2) "On a scale of 1–5, how closely did this problem match things you are interested in?" These items were only given for approximately one-third of the problems because students would tire of constantly answering these questions if they got it every time. The 1–5 ratings had a mean of 2.51 and a standard deviation of 1.26. A total of 51.5% of the responses were ratings of 1 or 2 (the median was 2 and the IQR was 2) suggesting the distribution was weighted towards the bottom of the scale. Appendix D shows how many rated problems, student raters, and ratings corresponded to the problems in terms of the number of prompts used to create the problems.

### Student open-ended feedback responses

The subset of students with parental consent received additional open-ended items after approximately 20% of the personalized problems (33 of 143 problems). The items were: (1) "Was the last problem interesting to you? Why or why not?", (2) "How do you think the last problem could be improved to better match with things you are interested in?" and (3) "Was there anything in the last problem that was incorrect relating to your interest area and how it works?".

## Data analysis

### Research questions 1–2

Each teacher prompt (RQ1) was segmented as the unit of analysis, with teacher verbal utterances considered alongside that segment, as well as the LLM's output (see Appendix C). Each prompt was coded for how teachers used ChatGPT. Prompts could be coded with multiple categories; for example, the prompt, "change the influencer's name to someone middle schoolers would know" was coded both with a category relating to swapping a popular culture interest and with a category relating to specifying the age of the target audience. Students' responses to the open-ended feedback questions (RQ2) were analyzed with one complete student written response as the unit of analysis. These responses were coded for themes that captured whether students responded positively or negatively to the problems, as well as feedback and suggestions students gave.

The analysis method was reflexive thematic analysis (Braun & Clarke, 2019a) with elements of codebook-based thematic analysis (Braun & Clarke, 2021), a flexible qualitative approach well suited to diverse data sources and grounded in the recognition that researchers play an active, interpretive role in the analytic process. Coding categories, themes, and the evolving codebook were developed organically through cycles of discussion, reflection, and refinement. Here the codebook was conceptualized as a dynamic record of the researchers' analytic thinking, which included code labels, examples, notes, and a coding framework to be used to organize the analysis. Broadly, the coding was informed by the theoretical framework – specifically, with issues of depth, grain size, and ownership (Walkington & Bernacki, 2014) being used as theoretical lenses. Guidelines for conducting reflexive thematic analysis emphasize that coding is flexible, organic and reflexive, and that it is necessarily influenced by the researchers who do the coding. For this reason, guidelines describe how the logic behind seeking to obtain measures of inter-rater reliability does not make sense in the context of this method (Braun & Clarke, 2019b, 2021).

A primary coder initially coded one half of the teacher prompting data (the data from Session 1) in an inductive, open-ended manner, writing an initial analytical memo and then comparing codes and themes identified with a second coder. An initial organizing framework for coding was proposed and applied to the dataset from Session 1 and shared with the larger research team for feedback and review. The primary coder also separately began coding the student feedback data from Session 1, writing another initial analytical memo and then discussing their coding with a second coder who had also looked for broad themes in the student feedback data. The second coder's role was to offer alternative interpretations and possibilities, checking the first coder's assumptions, rather than to independently reproduce the same coding scheme. After discussion, an initial coding scheme was proposed and applied to the data from Session 1. This coding was also presented to and discussed with the whole research team. At this point, another round of analytical memos were written, and the coding categories were refined through discussion to include a theoretical lens of depth, grain size, ownership, and ethical considerations, and to have coherence across both the teacher prompts and student feedback datasets. At this point, Session 1 and Session 2 data were coded together, and codes were refined as the additional data created new considerations and allowed the researchers to see places where the initial coding organization was working or was not working, and the coding was discussed with the larger research team again. After successive revisions to the analytical approach, the primary coder revisited the transcripts in their entirety and reconsidered previous coding organization based on developing interpretations. Table 1 shows an example of how raw responses were transformed over multiple iterations.

The primary researcher had a background in mathematics education, personalization theory, and generative AI in education, and had been present for many of the sessions with participating teachers. The second coder had a background in learning sciences and educational technology and had not been present for the data collection with teachers. The two other team members involved with the coding development were both former classroom mathematics teachers who were now in roles as mathematics education researchers, one focused on generative AI in mathematics teacher education and one focused on issues of equity in mathematics education. These two researchers had also directly worked with the participating teachers during the sessions and had worked extensively themselves with the dataset of teacher prompts and personalized problems.

## Research question 3

The number of prompts each teacher used for each problem was computed, along with the number of problems they posed, and the time it took them to pose each problem. For this analysis, there were 7 distinct teacher IDs, 494 distinct student IDs, 52 distinct problem IDs (one of the 53 problems that was rated lacked prompt data), and 3815 individual 1–5 ratings. Of the 3815 ratings, there were 6 cases where the dependent variable (student rating) was missing in the dataset and 145 cases where the number of prompts was missing due to missing chat log transcripts. There were instances when students simply chose not to start or continue with a problem set, as this would not have been recorded by the ASSISTments software. The correlation between interesting and closeness ratings was 0.87, indicating substantial conceptual overlap. The quantitative analyses presented here are correlational in nature, not allowing for the inference of causality.

**Table 1** Examples of how different organizing ideas were operationalized from participant responses

| Sample responses | Initial coding of session 1 data | Final coding of both sessions |
|---|---|---|
| "Can you make it about mac n cheese?"<br>"Give me a new twitch streamer."<br>"Use a brand name—i.e., Takis."<br>"Can we use Sprite, Coca Cola, Dr. Pepper, and Root Beer?" | In the initial data pass, a distinction was made between the first two where the teacher was swapping one reference for another, and the second two where the teacher introduced a reference without swapping it with an existing reference. This distinction was ultimately discarded. There was no attention paid to whether the LLM or teacher decided. The coding categories were: (1) Swap out a contextual reference and (2) Add specific pop culture reference | In the final data pass, the categories were ultimately collapsed between swap versus generate but were further separated based on whether the teacher or the LLM determined the exact reference. The categories were: (1) Add or swap specific popular culture reference (teacher decides) and (2) Add or swap particular popular culture reference (LLM decides). This was associated with the construct of "Grain Size + Ownership." |
| "What is an average price for sneakers at Foot Locker?"<br>"How large is San Diego County in miles?"<br>"What was the base model iPhone in 2023?"<br>"What is the name of the pond in [city] RI?" | In the initial data pass, these were coded into two different categories—the first two were in a category called 'Check Accuracy," and the latter two were in a category called 'Research." | In the final data pass, the categories were ultimately collapsed into the category "Brainstorming and Research." This was associated with the construct of "Depth." |

## Results

### Research question 1: what kinds of LLM prompts do teachers use to modify existing mathematics problems to be personalized to students' interests?

Research Question 1 involved analyzing the 143 *initial prompts* teachers used to first ask ChatGPT to pose a problem, as well as corresponding teacher verbal utterances. Moves for the initial prompt included giving ChatGPT the problem and asking it to either change the problem to be about a general popular culture area ("Can you make this problem about food?") or to add/swap a specific popular culture reference ("Can you rewrite this relating to Lululemon clothes?"). In the first case, the teacher is giving ChatGPT wide latitude to personalize to a broad topic and modify the problem in many possible ways. The teacher is not showing any specific knowledge of student interests other than knowing broad categories generally liked by adolescents. In the second case, the teacher is exhibiting more control over the content of the personalized problem, demonstrating knowledge about specific things their students like. These requests occur at different grain sizes and have different levels of ownership.

**Table 2** Categories of teacher prompts ($n = 415$ prompts)

| Construct | Category | Examples | Freq |
|---|---|---|---|
| Grain Size + Ownership (Teacher) | Give general popular culture topic area | “This is great, can we change the context to food.” | 19 |
| | Add or swap specific popular culture reference (teacher decides) | “Great. How about we name the video game Mario Kart.” | 105 |
| | Add or swap particular popular culture reference (LLM decides) | “Change bags of chips to a brand name.” | 56 |
| Depth | Realism of the context | “Please change the food items to typical concession stand snacks.” | 64 |
| | Realism of the numbers and/or quantities | “You cannot have half a coin. Can you rewrite this?”<br>“Can you change the price of the cleats to $70.” | 76 |
| | Brainstorming and research | “Can you tell me of a large outdoor event that teens would attend during the day?” | 45 |
| Depth + Grain Size | Tailor to age group | “Do it again but with a brand that is popular with middle schoolers.” | 37 |
| | Tailor to geographic location | “Could you make it about a college player from Texas?” | 30 |
| Grain Size + Ownership (Students) | Tailor to student names | “Change the names to names that sound like gen-z names.”<br>“It was [interesting] because it had my name in it.” | 17 |
| Ethical Issues | Readability issues | “Make this less wordy.”<br>“Can you combine the first two sentences.” | 27 |
| | Math difficulty and clarity | “Do it again, but use decimals like the original problem.” | 16 |
| | Output inappropriate or biased | “Change the MrBeast to a different Gen z influencer with no PR issues.” | 6 |
| Other | Other | “Don’t solve the problem, just rewrite it.” | 17 |

Note: Categories are not mutually exclusive

After the initial prompt, teachers made a total of 415 additional prompts. Table 2 shows how these prompts were distributed across categories; note this is a qualitative analysis, so the numbers are provided for transparency only. In some cases, the teacher gave the LLM a general topic area prompt, while in other cases the teacher gave a more specific popular culture reference. Sometimes, the *teacher* would indicate a specific popular culture reference, like "Change cooler ranch Doritos to flaming hot Cheetos." This distinction relates to grain size and ownership; the teacher's prompt suggests specific knowledge of student interests and involves more teacher control. This contrasts with cases where the teacher would want a specific popular culture reference added or swapped but would let the *LLM* decide. For example, one teacher requested "Use a brand's Instagram, but choose a brand popular with 12–14-year-olds."

Table 2 also shows that teachers used prompts to correct realism issues, which relates to depth. Sometimes teachers were concerned that the context of the problem was not something that students would be familiar with from their lives. One teacher asked, "Can we change the items they are buying to items that 7th grade students would buy?" when the LLM-generated problem included buying curtain rods; ChatGPT responded by making the problem about posters. A second category related to depth was when teachers felt the numbers and quantities in the story did not make sense. One teacher asked, "Can we make the driver obey the speed limit?" in a story originally about a plane modified to being about a car. The teachers would also sometimes use ChatGPT for brainstorming or research, including to check for realism. For example, one teacher asked, "Does esports have an app?" while another asked "How much does a YouTube premium account cost?" This shows the teacher trying to gain more familiarity with their students' interests, attending to problem depth, while also recognizing and attempting to address gaps in their contextual knowledge.

The next group of categories involved teachers asking ChatGPT to tailor problems to their students' age group, geographic location, or typical names. These moves could improve depth by allowing the LLM to better connect students' actual experiences in familiar locations, they could narrow grain size by matching students to content that is highly customized, and they could increase ownership as students potentially see themselves (e.g., their name) in problems.

There were also teacher prompts that corresponded to other issues with ethical dimensions related to developmental appropriateness and mathematical accuracy. In some instances, the LLM was prompted to fix the mathematical structures of the problem and in other instances, the general readability of the problem. For example, one teacher asked, "Change $t$ to $x$ in the problem" while another asked "Can you make it a multiplication equation?" There were also instances of ChatGPT generating content that teachers felt was inappropriate or biased, showing important instances where teachers may have been engaging in ethical monitoring. These included problematic celebrities being mentioned, a problem being modified by ChatGPT from being about one boy and one girl to being about two boys, and a scenario tracking a person's weight.

## Research question 2: what reasons do students give for finding personalized problems interesting or uninteresting, and what suggestions do they have for improving the problems?

There were 454 responses to the first student open-ended feedback, 449 to the second question, and 446 to the third question. For the first question, students gave 163 open-ended text responses implying the problem was interesting and 243 responses implying it was not, with the rest being uninformative or answers to a different question. The reasons they specified are in Fig. 1. Student response categories included giving no reason for their judgment and citing mathematical issues with the problem that were not directly relevant to the personalized context ("No because i don't like math."). Students also said they found the problem interesting because they liked its general popular culture topic ("Sure because food was in the problem.") or said they did not find the problem interesting because they disliked its general popular culture topic ("No I don't really like food or sports"). Similarly, students said they found a problem interesting because they liked its specific popular culture reference ("Yes it was because I listen to music (specifically on Spotify) a lot.") or that they did not find a problem interesting because they did not like its specific popular culture reference ("It was not that interesting to me because I don't play Roblox."). These responses suggest that students are considering the grain size of the problem. There were few instances of students mentioning that the names in the problems (which may allow ownership while indicating a smaller grain size) or the geographical locations of the problems (which may allow depth while indicating a smaller grain size) facilitated their interest. When they were giving feedback on the problems, there was little evidence that students explicitly attended to the dimension of ownership, especially compared to the degree to which they attended to the dimension of grain size.

The results from the second open-ended feedback question show 103 responses where students suggested that the problem did not need to be changed and 30 suggested that it should change but gave no suggestion. One requested change was to add or swap out a specific popular culture reference ("Switch TikTok with amazon" or "It could be about cats or like cute harp seals."). Other requests included changing the general topic ("Make it a sports question."). This suggests that students may have wanted finer-grained changes made to the problems rather than different broad interest categories, relating to grain size considerations.

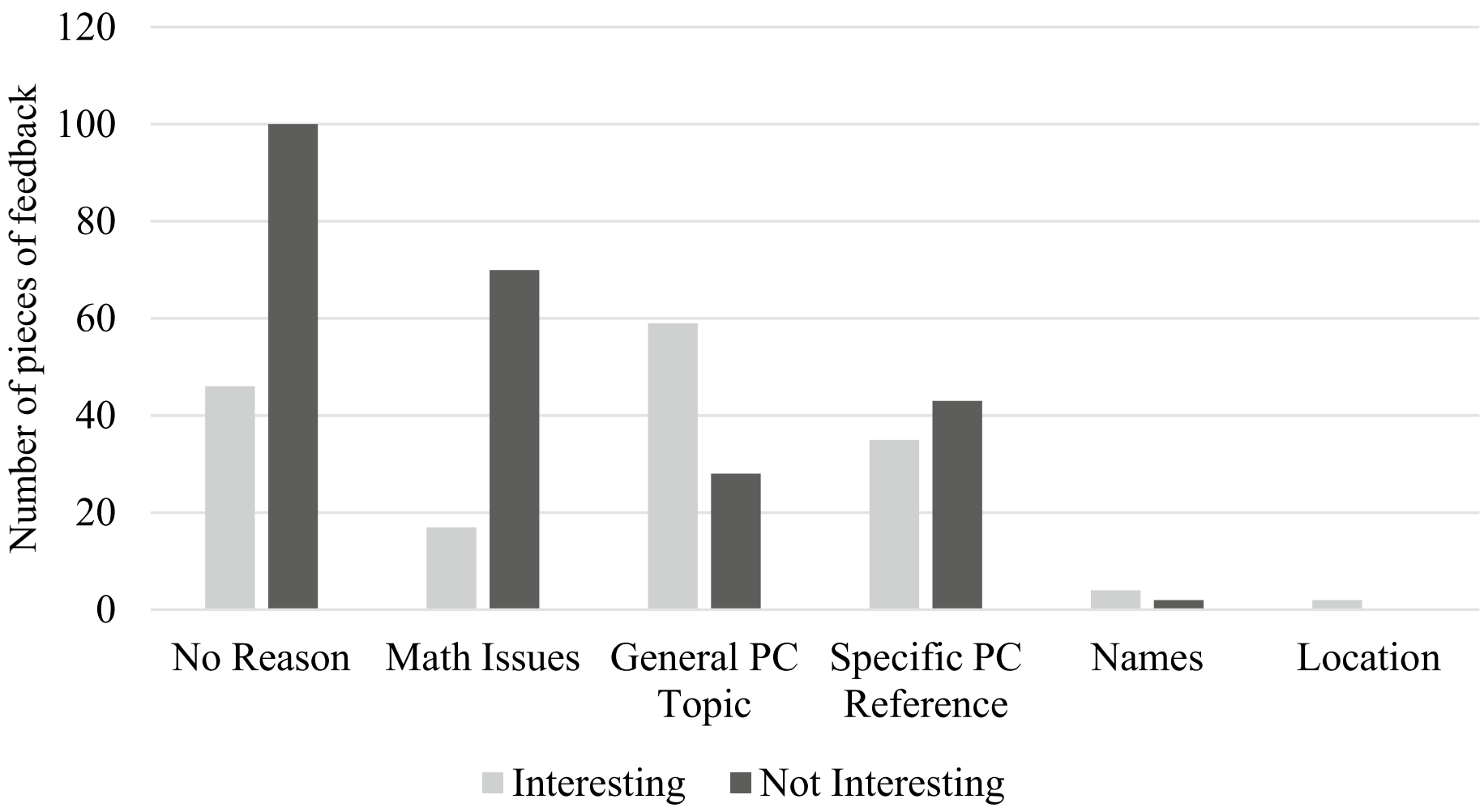

**Fig. 1** Reasons why students found problems interesting or not interesting. Frequencies summarize the number of written responses assigned each code within this dataset. They are presented descriptively, without inferential comparisons, and are not interpreted as measures of thematic importance

The final open-ended question related to whether anything in the problem was inaccurate with respect to students' interest areas, focusing on realism – it was trying to get the students to think about problem depth. In most cases, students said there was nothing inaccurate or did not answer the question because they said the problem was not in their interest area. There were cases where the student said something was inaccurate, but for the majority of these the student gave no information about what was inaccurate. In 13 cases, the student said some element of the context was inaccurate or not something they would ever do ("I wouldn't buy earbuds, but TikTok shop is really cool."). And in 5 cases students mentioned the numbers or quantities in the problem as being inaccurate (e.g., "Headphones aren't usually that much on TikTok shop."). Overall, students did not seem as attuned to the depth of the problems when giving feedback, focusing more on surface-level characteristics like popular culture references.

## Research question 3: how efficient is personalizing problems for teachers, and how does prompting relate to how the students rate those problems?

To write the 143 problems with ChatGPT, teachers used 558 prompts, for an average of 3.9 prompts per problem (*SD* = 2.6). The average problem took 246 s or approximately 4 min (*SD* = 165.7 s) to write. The teacher averages ranged from 2.0 to 5.5 min per problem, and from 2.7–6.6 prompts per problem. Analyses examined whether teachers using more prompts was associated with how students rated the problem. The average rating for the interestingness question was 2.46 (*SD* = 1.24) and for the closeness question was 2.53 (*SD* = 1.28). When examining the data visually, there are relatively small differences (Fig. 2). However, in both graphs, student ratings seem to initially peak around 3–5 prompts.

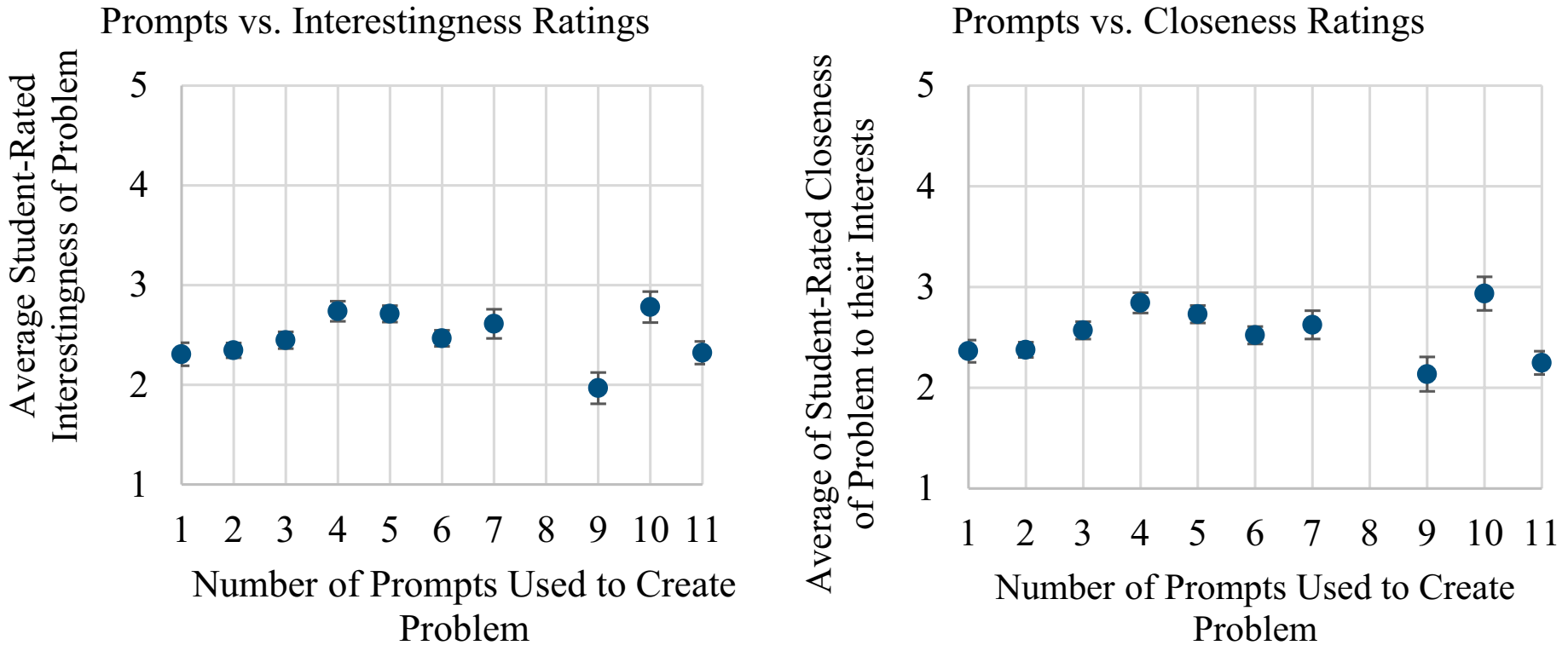

**Fig. 2** The relationship between how many prompts the teacher used, and student ratings of interestingness or closeness. Error bars represent standard error of the mean, calculated in terms of number of student raters (see Appendix D)

## Discussion

### Distributed personalization as a design process

The qualitative analysis for RQ1 showed that teachers made decisions about the depth, grain size, and ownership of the personalization (Walkington & Bernacki, 2014) as they engaged with ChatGPT. With respect to grain size, teachers would often either add or ask ChatGPT to add specific popular culture references—particular brands, celebrities, places, events, games, social media platforms, etc. – that they thought their students would be familiar with. Sometimes, especially for their initial prompt, they would use a broad grain size of a generic popular culture topic like "sports" or "music." These moves are consistent with research suggesting that smaller grain-size approaches to context personalization are more effective (Lin et al., 2024), and with research suggesting that teachers are concerned that LLMs might personalize at a grain size that is too broad (Walkington & Bainbridge, 2026). It also extends work suggesting that teachers vary in the degree to which they rely on their own professional knowledge versus data from students versus other sources when making connections to student interests (Rutherford et al., 2025)—introducing LLM agents as another source. These considerations may involve the teacher engaging in epistemic monitoring and demonstrating different epistemic stances—where they accept AI interest recommendations uncritically, versus consider these recommendations against what they have learned about their students.

In terms of depth, teachers were attuned to monitoring both the contextual and numerical realism of the problems, indicating some attunement to depth. Teachers wanted the problems to represent things that would actually happen in students' lives and wanted the numbers and quantities to be realistic. This is consistent with other research that suggests that some personalized scenarios can incorporate students' interests in only a shallow manner (Beauchamp & Walkington, 2024), that this is an issue in general with "school math" tasks (Palm, 2008), and that LLMs have issues with coming up with authentic and appropriate mathematical tasks situated in real-world scenarios (Einarsson et al., 2024; Sawyer & Aga, 2024; Srivatsa & Kochmar, 2024). There were not a lot of issues with the math problems themselves being unclear or ambiguous, as was seen in prior work on teacher problem-posing with LLMs (Biton & Segal, 2025; Leavy & Hourigan, 2020). This may be because the teachers were reformulating given problems rather than creating new problems, although results are exploratory given the small sample size. This builds on prior work exploring how teachers use LLMs to pose real-world problems (Biton & Segal, 2025), and work suggesting that teachers vary in terms of the depth at which they integrate students' interests when using LLMs (Rutherford et al., 2025). This also builds on work on epistemic agency (Wu et al., 2025), showing how teachers were focused on critically evaluating AI output from a uniquely human perspective—i.e., how realistic things were in the real world that they inhabit—using epistemic and ethical monitoring.

Finally, in terms of ownership, teachers varied in the degree they gave the LLM freedom to come up with popular culture references versus the degree to which they made these connections themselves based on their professional knowledge of their students. Sometimes, the initial freedom given to the LLM may have later been limited by the teacher in follow-up prompts, thus the degree to which the "broad topic" cases capture teachers' lack of control may vary. However, some of these results are consistent with research suggesting that some teachers may struggle to make meaningful connections between school mathematics and students' everyday activities (Chazan, 1999; Gainsburg, 2008) and that lack of familiarity with students' interests can be a barrier for teachers (Rutherford et al., 2025). It also builds on work about epistemic agency when using generative AI (Wu et al., 2025), raising the possibility that teachers may choose to be less critical of AI outputs and engage in less epistemic monitoring in cases where they lack deep knowledge of their students' interests, and perceive AI as potentially having more actionable knowledge than they do.

While teachers can play a key role in the AI-mediated personalization process by engaging in ethical and epistemic monitoring, it is important to acknowledge that placing a teacher in-the-loop does not automatically ensure that these behaviors will take place. Teachers need to understand the key knowledge-related criteria to judge AI outputs with, which includes having appropriate mathematical and pedagogical knowledge. They also must have the time, space, and agency to notice and act upon these criteria, and be working with AI production pipelines that allow for iterative revision and critical reflection on AI outputs. In addition, even when monitoring does occur, there is no guarantee that it will be effective.

## Scalability and grain-size tradeoffs

From the qualitative analysis of the student feedback, results suggested that students in the sample often asked for specific, small-grain-size connections to their popular culture interests, and disliked specific references when they do not match with their personal preferences. So, for example, two students might like music, but one might hate Taylor Swift-themed questions, while the other might want only Taylor Swift-themed questions., This issue of grain size is difficult to reconcile while there is a teacher-in-the-loop. Smaller grain sizes are more feasible with systems where the LLM faces the individual student (e.g., Khan Academy, n.d.), or where the student interacts with the LLM to generate the problems (Norberg et al., 2025).

In terms of depth, there were only a few cases where the student was able to articulate something about the problem that was unrealistic. From this, a tempting conclusion might be that grain size is more important to students than depth, and that if a trade-off must be made, grain size should be prioritized. However, many mathematics educators want students to believe that mathematics is a sensible way to model the world (Palm, 2008; Vos, 2018) rather than arbitrary algorithms. From this perspective, depth is important to students adopting an important worldview about mathematics as a discipline in society, which can make future mathematics courses and endeavors more meaningful (Lampert, 1992).

Students rarely gave feedback related to their ownership of the personalized problems. It may be that the dimension of ownership was less salient to them than that of depth or grain size. Or it may be that the task of giving feedback tended to elicit student responses that assumed that their practice problem sets would adhere to more typical norms of school mathematics where students do not always get to exercise ownership of the problems they solve. There were occasional exceptions to this; one student described liking the personalized problem because "It shows my opinion," while another stated wanting to change their personalized problem, "I have some name suggestions for your problems to match my interest."

There were also important instances where the teachers acted in a role as an ethical and epistemic monitor of LLM outputs. Teachers monitored LLM content for issues with realism, readability, mathematical accuracy and consistency, and even occasionally safety and bias. These actions are interpreted as indicators of epistemic monitoring when there was evidence that teachers drew on their pedagogical or mathematical knowledge to judge an AI output, and then acted on that judgment. These episodes illustrate evaluative practices consistent with aspects of an evaluativist epistemic stance. If teachers are taken out-of-the-loop, the possibility of teachers having the opportunity to enact these ethical decisions and protections for their students, and to learn from these opportunities, may be lost.

## Exploratory design implications

The present data suggested that personalizing a single problem took teachers in the sample approximately 4 min. Given that a typical problem set has around 10 problems, a teacher would spend 40 min per assignment to create *one* alternate version of each original problem. The vision of this project was that teachers would be able to create 4 versions of each problem that corresponded to 4 different interests. Once the problem-posing sessions with teachers started, however, it was clear this was going to be untenable. Teacher-in-the-loop personalization may be structurally constrained to particular grain sizes due to time costs, independent of pedagogical preferences of teachers and students, but more research is needed to determine this. This is important when considered alongside other studies showing that teachers worry about the time of creating lessons and tasks with LLMs (e.g., Kuusemets et al., 2024; Rutherford et al., 2025). It should become a norm in the research literature for all studies to report the amount of time teachers spent interacting with the LLM. This time could be viewed as a pedagogical investment—especially in cases where the teacher uses the time to learn more about students' interests, ways to effectively pose math problems, and using LLMs to enhance their instruction.

When looking at descriptive data on prompting efficiency, exploratory trends suggested there may be a pattern where using a modest number of prompts is associated with the most favorable student feedback. One interpretation of this pattern, informed by the qualitative findings for RQ1, is that engaging in follow-up prompts gave teachers opportunities to address issues with realism, fit to students' interests, readability, and mathematical structure. Thus, problems created with very little iteration may sometimes retain such weaknesses. This relates to research suggesting that improvement requires structured iteration (Ericsson et al., 1993), and that without this iteration teachers may show reliance on weakly-contextualized AI outputs. The results also suggested too many prompts could signal something was going wrong with the personalization. Following research on human–computer interaction, repeated iterative refinements may only lead to cyclic improvement until gains plateau (Nielsen, 1994).

One important alternative explanation for the findings is that teachers needed more prompts to personalize problems that were more difficult or complex—and that students tended to have greater dislike for more difficult problems. This could partially explain why using many prompts was sometimes associated with lower student ratings, but not why using few prompts was associated with lower student ratings. In addition, teachers' familiarity with the mathematical content and interest context and the quality of the initial AI output may influence both the amount of prompting and students' subsequent ratings. Future research should examine the degree to which human–AI instructional design operates within a bounded iteration window that takes into account the sometimes competing issues of shallow acceptance of GenAI outputs versus over-optimization where teachers go back-and-forth over and over again. Experimental research is needed to explore these trends, while also collecting qualitative data that determines when spending time or prompts developing a problem is a pedagogical investment versus an inefficient use of time.

These findings may extend other work suggesting that teachers may benefit from learning more about effective prompting strategies (Dilling & Herrmann, 2024). It also gives a starting point when considering that in some situations a teacher might be using too many prompts and may be excessively struggling. As an example, future teacher-facing LLM systems may be designed to move to special protocols when teachers engage in specific types of prompting behaviors when designing a single task, perhaps telling them that their request is unlikely to work (e.g., some problems are not well-suited to be personalized) or proactively giving them explicit suggestions on what to try next, rather than reactively responding. This could address some of the critiques from teachers that using LLMs is time-consuming (Kuusemets et al., 2024). But ultimately, more research is needed on when and how this might occur.

## Significance and conclusion

A limitation of this study is that it only examined proximal outcomes of student interest ratings and open-ended feedback, rather than students' performance and learning. Students' ratings of the problems may not reflect their performance or learning from solving the problems. However, student interest is a theoretically meaningful proximal indicator in personalization research (Bernacki & Walkington, 2018; Walkington & Bernacki, 2019). This study was designed to be exploratory, so it did not have a control group that would allow for the comparative analysis of performance outcomes; this is an important direction for future work. Limitations also included that teachers were being observed while prompting, which may have shaped prompting behaviors in ways that differ from naturalistic classroom planning contexts. Finally, only seven teachers participated, all of whom taught seventh-grade mathematics and used IM through ASSISTments. Although each teacher contributed multiple problems and engaged in two design sessions, these repeated episodes do not substitute for a broader teacher sample. In addition, the teachers engaged in specific training and reflection sessions which would not be the case for a general population of teachers engaging in personalized task creation. Accordingly, the findings should be interpreted as descriptive and hypothesis-generating within this bounded implementation, rather than as estimates of how frequently these practices occur among teachers or as evidence that the observed patterns will transfer to other grades, curricula, platforms, AI models, or national contexts.

This study extends Walkington and Bernacki's (2014) framework for personalization, contributing some exploratory design considerations for AI-supported personalization. The design considerations for future study include: (1) operationalizing depth, grain size, and ownership in LLM-mediated personalization design, (2) identifying how these constructs are manifested in teacher prompts and student feedback, (3) illustrating situated instances of teachers' epistemic monitoring of mathematical and contextual representations, alongside ethical monitoring of content considered inappropriate or biased, (4) exploring the relationship between efficiency and time measures and student feedback, and (5) revealing potential structural grain-size limitations with teacher-in-the-loop designs.

Future research should consider teachers' iterations with GenAI when crafting learning tasks to determine whether the volume of prompts represents an important area when considering human-AI instructional co-design. When teachers engage in minimal iterative prompting when personalizing math tasks, this may indicate surface-level acceptance of AI output without sufficient refinement for depth or grain size alignment. In contrast, extended prompting cycles could signal over-optimization, potentially leading to drift from the original mathematical structure or unnecessary complexity in the narrative context. The study also shows how it is important that teachers be able to enact epistemic agency by evaluating and revising AI outputs against professional criteria. Merely having teachers in-the-loop does not guarantee that teachers will activate their epistemic agency and engage in epistemic monitoring, but it does allow for these behaviors to arise.

Within this implementation, the findings suggest a potential scalability tradeoff between teacher control and students' preference for fine-grained interest matching that has not been documented in prior personalization research. The process of personalizing content with GenAI even at a broad grain size in this study represented a significant investment of teachers' planning time. However, other research suggests that by having a teacher in-the-loop, teachers may gain opportunities to engage in professional learning (Walkington et al., 2026b). They may learn about their students' interests, how these interests work quantitatively, and how to create a classroom culture around the valuing of student assets, that would not be possible if personalization was completely offloaded to GenAI. Conversely, it has been documented that students worry that allowing teacher functions to be taken on by GenAI will create distance between students and teachers (Center for Democracy & Technology, 2025).

Collectively, these findings explore accounts of teachers using AI and allow for important directions for future research and design to be uncovered. From this sample, this study provides an exploratory account of heuristics for building AI personalization systems that support effective teacher prompting, scaffolding depth without sacrificing efficiency, and preserving meaningful teacher ownership while addressing students' desire for fine-grained personalization. In doing so, the study advances a design-oriented research agenda for generative AI in education that balances scalability, pedagogical integrity, and teacher professionalism.

**Supplementary Information** The online version contains supplementary material available at https://doi.org/10.1007/s11423-026-10756-2.

**Acknowledgments** This work was supported by the National Science Foundation under Grant DRL 2341948. Any opinions, findings, conclusions, or recommendations expressed in this material are those of the author(s) and do not necessarily reflect the views of the National Science Foundation.

**Funding** Open access funding provided by SCELC, Statewide California Electronic Library Consortium. Funding was provided by Directorate for STEM Education, DRL 2341948, Candace Walkington.

## Declarations

**Competing Interests** The authors declare that they have no known competing financial interests or personal relationships that could have appeared to influence the work reported in this paper. The authors declare the following financial interests/personal relationships which may be considered as potential competing interests: Candace Walkington reports financial support was provided by National Science Foundation. If there are other authors, they declare that they have no known competing financial interests or personal relationships that could have appeared to influence the work reported in this paper.

**Publisher's Note** Springer Nature remains neutral with regard to jurisdictional claims in published maps and institutional affiliations.

**Candace Walkington** is a professor of mathematics education and learning sciences at Southern Methodist University.

**Mingyu Feng** is the research director of the learning and technology division at WestEd.

**Tiffini Pruitt-Britton** is a researcher in educational equity and mathematics education at American Institutes for Research.

**Theodora Beauchamp** is a Ph.D. student in mathematics education at Southern Methodist University.

**Andrew Lan** is an associate professor of computer science at University of Massachusetts Amherst.